%% file: main.tex
\documentclass[conference]{IEEEtran}
\IEEEoverridecommandlockouts
\usepackage{cite}
\usepackage{amsmath,amssymb,amsfonts}
\usepackage{algorithmic}
\usepackage{graphicx}
\usepackage{textcomp}
\usepackage{xcolor}
\usepackage{makecell}
\usepackage{url}
\usepackage{xspace}        

\usepackage[a4paper, total={184mm,239mm}]{geometry}
\def\BibTeX{{\rm B\kern-.05em{\sc i\kern-.025em b}\kern-.08em
    T\kern-.1667em\lower.7ex\hbox{E}\kern-.125emX}}
\begin{document}

\newcommand{\NAME}{OMU\xspace}

\title{\NAME: A Probabilistic 3D Occupancy Mapping Accelerator for Real-time OctoMap at the Edge}

\author{\IEEEauthorblockN{Tianyu Jia$^2$, En-Yu Yang$^1$, Yu-Shun Hsiao$^1$, Jonathan Cruz$^1$, David Brooks$^1$, Gu-Yeon Wei$^1$, Vijay Janapa Reddi$^1$}
\IEEEauthorblockA{$^1$Harvard University, Cambridge, MA, $^2$Peking University, Beijing, China}
}

\maketitle

\input{0_Abstract}
\input{1_Introduction}

\input{2_Background}

\input{3_Workload}

\input{4_Hardware}

\input{5_Methology}

\input{6_Experiments}

\input{7_Conclusion}


\section*{Acknowledgment}

This work is supported in part by JUMP ADA Research Center, DARPA DSSoC program, and the NSF Awards 1718160 and 1955422.

\bibliographystyle{IEEEtran}
\bibliography{main}

\end{document}

%% file: 0_Abstract.tex
\begin{abstract}
Autonomous machines (e.g., vehicles, mobile robots, drones) require sophisticated 3D mapping to perceive the dynamic environment. However, maintaining a real-time 3D map is expensive both in terms of compute and memory requirements, especially for resource-constrained edge machines. Probabilistic OctoMap is a reliable and memory-efficient 3D dense map model to represent the full environment, with dynamic voxel node pruning and expansion capacity. This paper presents the first efficient accelerator solution, i.e. \NAME, to enable real-time probabilistic 3D mapping at the edge. To improve the performance, the input map voxels are updated via parallel PE units for data parallelism. Within each PE, the voxels are stored using a specially developed data structure in parallel memory banks. In addition, a pruning address manager is designed within each PE unit to reuse the pruned memory addresses. The proposed 3D mapping accelerator is implemented and evaluated using a commercial 12~nm technology. Compared to the ARM Cortex-A57 CPU in the Nvidia Jetson TX2 platform, the proposed accelerator achieves up to 62$\times$ performance and 708$\times$ energy efficiency improvement. Furthermore, the accelerator provides 63~FPS throughput, more than 2$\times$ higher than a real-time requirement, enabling real-time perception for 3D mapping.

\end{abstract}


%% file: 1_Introduction.tex
\section{Introduction}

3D Mapping is an essential perception process in autonomous machines to build polygonal representation for the environment. With the information of the 3D map, autonomous machines can perceive the surrounding environment and perform several safety-critical autonomous execution tasks, such as localization, motion planning, and collision detection \cite{mp32021cvpr}.  

\begin{figure}[!t]
    \centering
    \includegraphics[width=1\columnwidth]{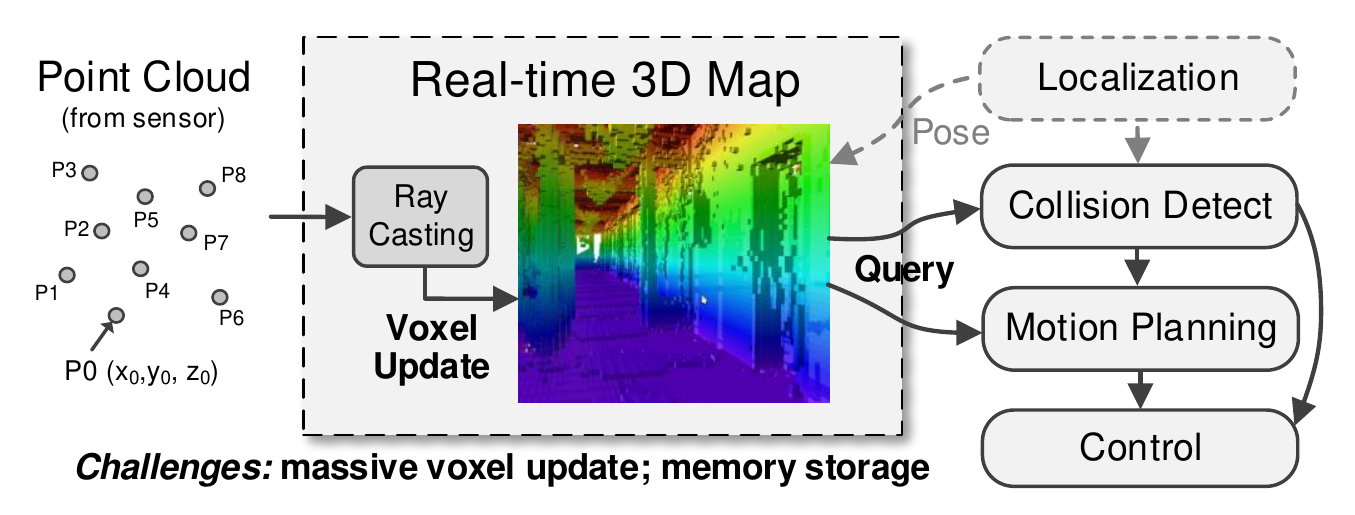}
    \vspace{-20pt}
    \caption{The computation pipeline in a autonomous machine. 3D Mapping is a crucial task that serves many (safety) purposes and can consume as much as 70\% the execution time~\cite{Behzad2018mavbench}.}
    \label{fig:overall_flow}
    \vspace{-10pt}
\end{figure}

To build a 3D map, the sensor data generated from sensors, such as the 3D point cloud, is streamed into the computation pipeline to create and update the map, such as a corridor 3D map example in Fig. \ref{fig:overall_flow}.
Within the end-to-end computation of autonomous systems, the perception stage (i.e., parse sensor data, build the 3D map, and localization) is a computationally intensive process for tracking real-time changes in the environment. 
For example, in a package delivery task for a micro aerial vehicle (MAV), the 3D map generation can take more than 70\% of the total runtime for autonomous navigation \cite{Behzad2018mavbench}. 

Maintaining and querying a 3D map in real-time is costly in terms of both memory and computation resources. As many 3D maps discretize the environment into cubic volumes (i.e., voxels), there is high memory access and capacity requirement for 3D map build and update. Due to the randomness of the input map voxel coordinates, the memory access pattern is highly irregular. The bandwidth of the memory is the main performance bottleneck impeding the real-time 3D mapping.
Furthermore, the 3D map also requires large memory storage, which further increases with a higher resolution (i.e., smaller voxel size). For example, in a 3D scan dataset \cite{octomap_dataset}, a full 3D map with a resolution of 10~cm could require significant memory storage ranging from 10~MB to more than 5~GB \cite{hornung2013octomap}. 

In addition to the high memory access demand, the large volume of sensor data leads to a significant computation challenge for real-time 3D mapping. 
For example, the Microsoft Kinect sensor produces 9.2 million 3D points per second \cite{Biswas2014icra}.
Such a high volume of points leads to low hardware throughput, i.e., below the real-time requirement 30 frames per second (FPS) \cite{singh2018cvpr}. 
Prior work explored the sampled frames or sparse points cloud for 3D maps, but most real-time 3D mappings still need to run on high-end CPUs \cite{hess2016icra} or GPUs \cite{gpuvoxel2014iros}, leading to significant power and form factor overhead.
In resource and energy-constrained autonomous machines at the edge, enabling real-time 3D mapping is even more challenging. 


\begin{figure}[!t]
    \centering
    \includegraphics[width=0.95\columnwidth]{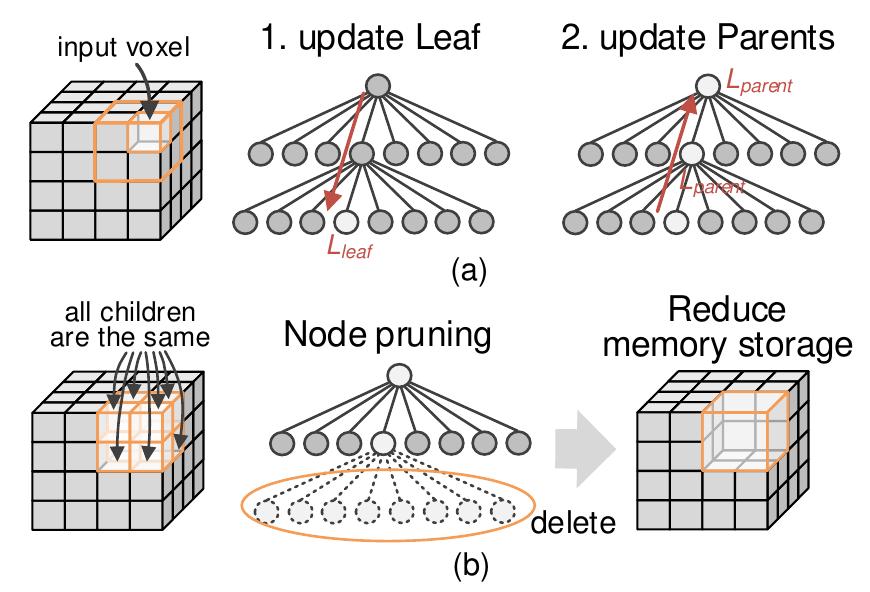}
    \vspace{-10pt}
    \caption{(a) The probability update operations for every voxel input, (b) the pruning operation when all the children have the same occupancy probability (tree depth=2 for illustration).}
    \label{fig:octree_steps}
    \vspace{-10pt}
\end{figure}

In this work, we present the first 3D mapping accelerator for an efficient and widely used 3D occupancy mapping algorithm, i.e., OctoMap \cite{hornung2013octomap}. OctoMap is one of the most widely used approaches to represent the entire 3D environment due to its memory-efficient ``Octree'' structure; it is akin to the H.264 codec for video compression. In the OctoMap, the 3D map voxels are stored in the Octree, where all nodes are recursively divided into octants (i.e., eight children), as shown in Fig. \ref{fig:octree_steps}. The occupancy of each voxel is represented by probability, forming a reliable 3D map. Furthermore, the map leaf nodes can be pruned (compressed) based on the likelihood of children nodes. Although OctoMap is a state-of-the-art approach with promising results to save memory storage, it can still not meet the real-time throughput requirement on most desktop CPUs and edge device CPUs (based on our analysis in Section \ref{sec:workload_analysis}). Therefore, we develop an efficient 3D OctoMap accelerator, i.e. OctoMap Processing Unit (OMU), in this work to enable its operation in real-time for edge machines. 

To improve the hardware accelerator's performance, we first analyze the performance bottlenecks of OctoMap on CPUs. We observe that the voxel node pruning and expansion consume most of the runtime. Based on this workload analysis, we designed \NAME\ with three key features. First, we update the map's voxels in parallel processing element (PE) units to maximize the compute throughput by up to 8$\times$. Second, we designed a special data storage structure and parallel memory banks to improve voxel store/fetch throughput by another 8$\times$. Third, we created a pruning address manager to manage the pruned addresses and maintains high memory utilization.

We evaluated the proposed \NAME\ accelerator after synthesis, place and route using a commercial 12~nm CMOS technology. Compared to a state-of-the-art desktop Intel i9 CPU and a more conventional robotics platform (i.e., ARM Cortex-A57 CPU in Jetson TX2), the accelerator achieves 13$\times$ and 62$\times$ performance improvement, respectively, and a 708$\times$ energy efficiency improvement compared to the ARM A57 CPU. \NAME\ only consumes 250.8~mW of power and achieves 63~FPS in throughput, thus enabling low-power real-time perception in resource-constrained autonomous machines. 



The contributions of our work are as follows: 
\begin{itemize}
    \item We present a comprehensive workload breakdown and characterize the bottlenecks in 3D OctoMap. We observe that the node pruning stage consumes the most run-time, which we optimize via parallel access of children nodes.
    \item We develop a series of hardware acceleration techniques for probabilistic 3D mapping, including a method for parallel map-voxel updates, an efficient memory data storage structure, and a dynamic pruning-address management.
    \item We demonstrate a 12~nm probabilistic 3D mapping accelerator \NAME. The experimental results show significant performance and power benefits, and 2$\times$ higher throughput than the real-time requirement for 3D mapping tasks.
\end{itemize}

%% file: 2_Background.tex
\section{Background}

Over the past decades, many space representation approaches have been studied to map the unstructured environments into a digitized representation.  The 3D maps contain the environment information in three dimensions, which is demanded in the perception of autonomous machines. 

\textbf{Dense vs. Sparse Map:} 
Depending on whether the map can represent the whole 3D environment, the map models can be categorized as either dense or sparse maps.
For navigation and manipulation tasks in autonomous machines or robots, a ``dense'' 3D map model, which can cover the entire environment space, is desired for collision detection and motion planning. Several dense 3D models have been widely used to represent the environments, such as point clouds, elevation maps, multi-level surface maps, Octree \cite{hornung2013octomap}, and signed distance fields \cite{voxblox2017iros}. However, only Octree and signed distance fields can represent the unknown space. 
Compared to the dense map, sparse 3D maps are built based on extracted features or landmarks \cite{2018vlsi_navion}\cite{2019isscc_slam}\cite{Artal2015orb-slam}. However, these sparse 3D maps cannot represent the unknown space, which is necessary for the collision detection task in autonomous motion planning. 



\textbf{Probabilistic Representation:} To represent the space, each space point or voxel can be simply recorded in a binary manner as either occupied or free. However, given the uncertainty of sensor measurement (e.g. noise, errors), the simple binary space representation often leads to an inaccurate map ~\cite{hornung2013octomap}. 
A more reliable probabilistic map model can represent the map data by its occupancy probability instead of only representing the map voxels by occupied or free. During the map update, the map voxel updates its occupancy probability on top of the prior probability. The map voxels can be merged (pruned) with their neighbors with the same probability to save memory storage with the occupancy probability values. Although the probabilistic map is reliable and memory efficient, it needs a unique data storage structure and additional hardware.

\textbf{Mapping Accelerators:} To accelerate the perception of autonomous systems, there are a few specialized accelerators that have been developed for resource-constrained autonomous robots or drones, as shown in Table \ref{tab:acc_hw}. For example, an OctoMap based motion planning accelerator was developed for collision detection in robots \cite{lian2018dadup}\cite{han2020daduseries}. However, the accelerator preloads and processes a lossy 3D map in batches, missing the challenges of real-time map building. There are also SLAM-based accelerators designed for real-time applications \cite{2018vlsi_navion}\cite{2019isscc_slam}, while they only build a sparse 3D map using extracted features and cannot represent the entire environment including the unknown space. 
Different from prior arts, in this work, \emph{we propose an accelerator solution---\NAME\ to support \textbf{dense} and \textbf{probabilistic} 3D mapping in \textbf{real-time}. }

\begin{table}[t!]
\centering
\caption{Comparison of Mapping Accelerators}
\vspace{-5pt}
\begin{tabular}{|l|c|c|c|c|c|}
\hline 
    & \makecell{Dadu-p\\\textbf{\cite{lian2018dadup}}}  & \makecell{Dadu-cd\\\textbf{\cite{han2020daduseries}}}  &   \makecell{Navion\\\textbf{\cite{2018vlsi_navion}}} &  \makecell{CNN-SLAM\\\textbf{\cite{2019isscc_slam}}}  & \makecell{This\\{Work}} \\ \hline
Dense Map    & \checkmark   & \checkmark  & No     & No    & \checkmark      \\ \hline
Probabilistic      &No   & No  &No   & No     &\checkmark        \\ \hline
Real-time     &No   & No  &\checkmark   & \checkmark     &\checkmark        \\ \hline
\end{tabular}
\label{tab:acc_hw}
\vspace{-15pt}
\end{table}%

%% file: 3_Workload.tex
\section{OctoMap Workload Analysis} \label{sec:workload_analysis}

To understand the computation bottleneck and guide the OctoMap accelerator design, we first introduce the basic operations in OctoMap and analyze its workload breakdown.

\subsection{OctoMap Overview}
As the map generation example illustrated in Fig. \ref{fig:overall_flow}, the 3D sensors generate a series of point clouds for 3D map building. A ray casting kernel will process the point cloud data to identify free voxels along the ray, while the point cloud will be registered as occupied voxels.
In the OctoMap, the 3D space is discretized into equal-sized voxels and stored in an Octree structure, where all nodes are recursively divided into 8 children, as shown in Fig. \ref{fig:octree_steps}. Each leaf node \emph{n} is represented by the log-odds notation of the occupancy probability, as shown in equation (1), where \text{P(n)} is a prior probability \cite{hornung2013octomap}.

\begin{equation}
    L_{\text{leaf}}(n) = \log\left[ \frac{P(n)}{1-P(n)}\right] 
\end{equation}

There are three basic operations to update the occupancy status of one voxel: \emph{update leaf node}, \emph{recursively update parent nodes}, and \emph{node pruning}. First, based on the coordinates of the input voxel, a corresponding leaf node is found with its occupancy probability updated. Due to the log-odds notation of the probability value, only a simple addition operation is needed to update its probability, as shown in equation (2). Next, to enable efficient memory storage and multi-resolution queries, the occupancy probabilities of the parents are recursively updated from the bottom to the root. The probability policy of the parent nodes takes the maximum occupancy of all its eight children, as illustrated by equation (3). If all the children have identical occupancy during the parent update, the children nodes are pruned while its parent becomes a leaf. On the contrary, if all the children no longer have the same occupancy, the pruned leaf node is expanded. Experiments show that Octree pruning can significantly reduce the memory storage by up to 44\% with no accuracy loss~\cite{hornung2013octomap}.

\vspace{-5pt}
\begin{equation}
    L_{\text{leaf}}(n|z_{1:t}) = L(n|z_{1:t-1}) + L(n|z_{t}) 
\end{equation}

\vspace{-10pt}
\begin{equation}
    L_{\text{parent}}(n) = \max_{i=8} \ {L_{\text{leaf}}(n_i)} 
\end{equation}

During the queries of the map, the occupancy value will be fetched based on the voxel coordinates. Based on the predefined clamping thresholds, the log-odds occupancy value can be classified as a status of occupied, free, or unknown. 

\subsection{Runtime Breakdown and Bottleneck Analysis}

\begin{table}[t!]
\centering
\caption{Details of OctoMap 3D scan dataset.}
\vspace{-5pt}
\scalebox{0.9}{%
\begin{tabular}{|l|l|l|l|}
\hline
 &   \textbf{FR-079 corridor}  &   \textbf{Freiburg campus}  &   \textbf{New College}   \\ \hline
Scan Number    & 66     & 81    &92361      \\ \hline
Average Points / Scan      &89x$10^{3}$   &248x$10^{3}$    &156        \\ \hline
Point Cloud (x$10^{6}$)        &   5.9     &  20.1    &   14.5            \\ \hline
Voxel Update (x$10^{6}$)   &   101         & 1031     &  449       \\ \hline
i9 CPU Latency (s)   &   16.8         & 177.7     &  77.3       \\ \hline
CPU Throughput (FPS)    &   5.23        & 5.03      & 5.04       \\ \hline
\end{tabular}
}
\label{tab:map_details}
\vspace{-10pt}
\end{table}%

To characterize 3D mapping as a computational task, we performed runtime analysis for a variety of map graphs from the OctoMap 3D scan dataset \cite{octomap_dataset}. Each map we select is a representative sample from the dataset suite. The 3D maps have been run on a beefy Intel i9-9940X desktop CPU, with the experiment details listed in Table \ref{tab:map_details}. They contain multiple 3D laser scans, with every scan providing different numbers of 3D points. In this experiment, all the maps use the same resolution (i.e., voxel size) of 0.2~m. For example, to complete the 3D mapping of the FR-079 corridor map, the 3D laser generates 5.9 million points in the point cloud, which leads to more than 100 million voxel updates. 
To build the entire 3D map, the i9 CPU takes between 16 seconds and 3 minutes. The CPU throughput is only 5 frames per second when equivalently derived for common 320x240 sensor image size, and is much lower than a 30~FPS requirement in real-time application \cite{singh2018cvpr}.

\begin{figure}[!t]
    \centering
    \includegraphics[width=0.85\columnwidth]{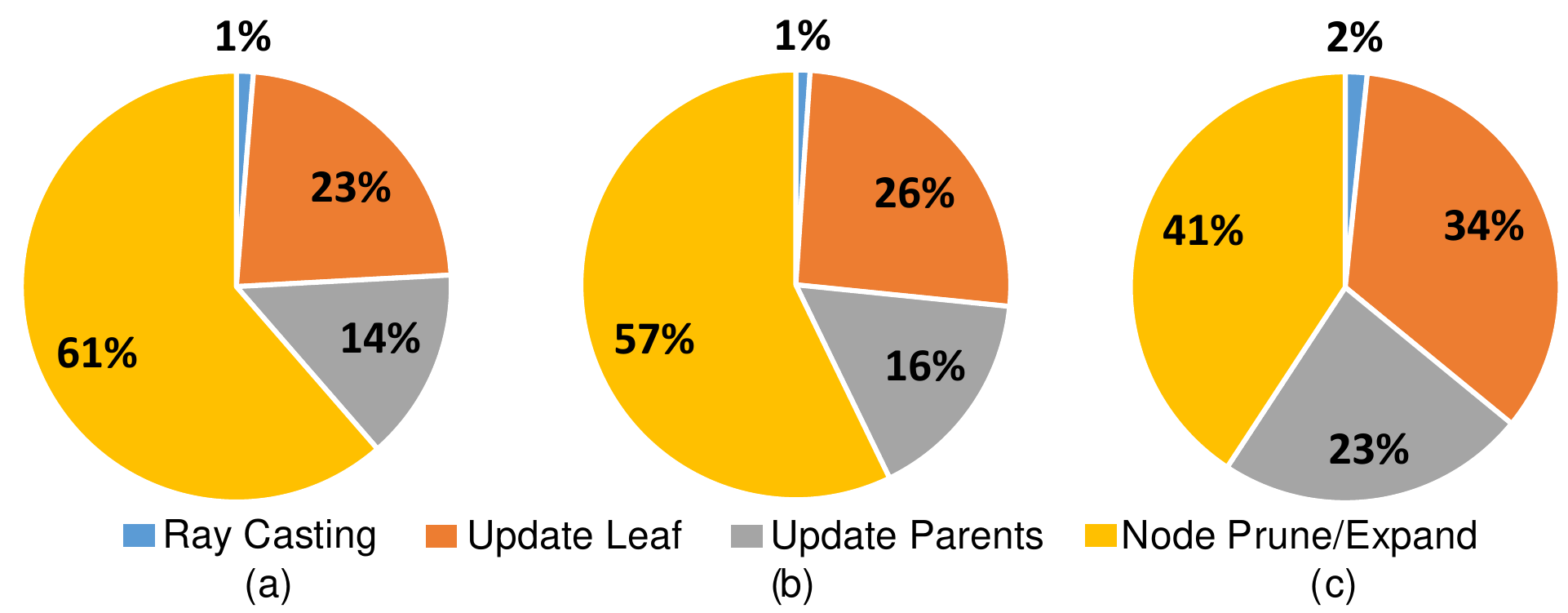}
    \vspace{-10pt}
    \caption{Runtime breakdown in OctoMap workloads (a) FR-079 corridor, (b) Freiburg campus, and (c) New College.}
    \label{fig:runtime_breakdown}
    \vspace{-20pt}
\end{figure}

To understand the hardware requirements of OctoMap, we break down the runtime of each OctoMap operation (i.e., ray casting, update leaf, update parents, and node prune or expand) as shown in Fig. \ref{fig:runtime_breakdown}. The node prune and expand stages consume most of the runtime in all of the maps and is the workload bottleneck. This bottleneck is due to the ample tree prune leading to significant irregular memory access for children nodes. The limited memory bandwidth degrades the children node access and becomes the performance bottleneck for OctoMap. The node prune in the New College dataset takes less total runtime than the other two maps, mainly due to the fewer points per scan leading to less node pruning.

It is worth to mention that the number of voxel updates can be reduced by voxel overlap search during ray casting. Hence the overlapped voxels only need to be updated once to save the memory access cost. However, to enable the voxel overlap search, the ray casting needs special voxel hashing and complex hardware acceleration, as \cite{kar2020raycasting}. In this work, we focus on the voxel map integration challenge, which can be combined with other advanced ray casting acceleration.

%% file: 4_Hardware.tex
\section{Hardware Acceleration for 3D OctoMap}

To improve the real-time performance of 3D OctoMap, we developed a series of acceleration techniques, including parallel Octree update (for 8$\times$ compute throughput), efficient data storage and parallel memory storage (for additional 8$\times$ memory bandwidth), and dynamic prune address management (for high memory utilization), as shown in Fig. \ref{fig:pe_uarch}.

\subsection{Parallel Octree Update}

In the conventional OctoMap update, the input voxels are updated in series during the mapping process, as there is potential update dependency between voxels. Even the latest version of OctoMap that is available on GitHub is still a single-threaded application library.\footnote{At the time of writing, the latest release of OctoMap is 1.9.6. It is publicly accessible online at \url{https://github.com/OctoMap/octomap}} This series voxel update significantly limits the throughput of the OctoMap computation. 

\begin{figure}[!t]
    \centering
    \includegraphics[width=1\columnwidth]{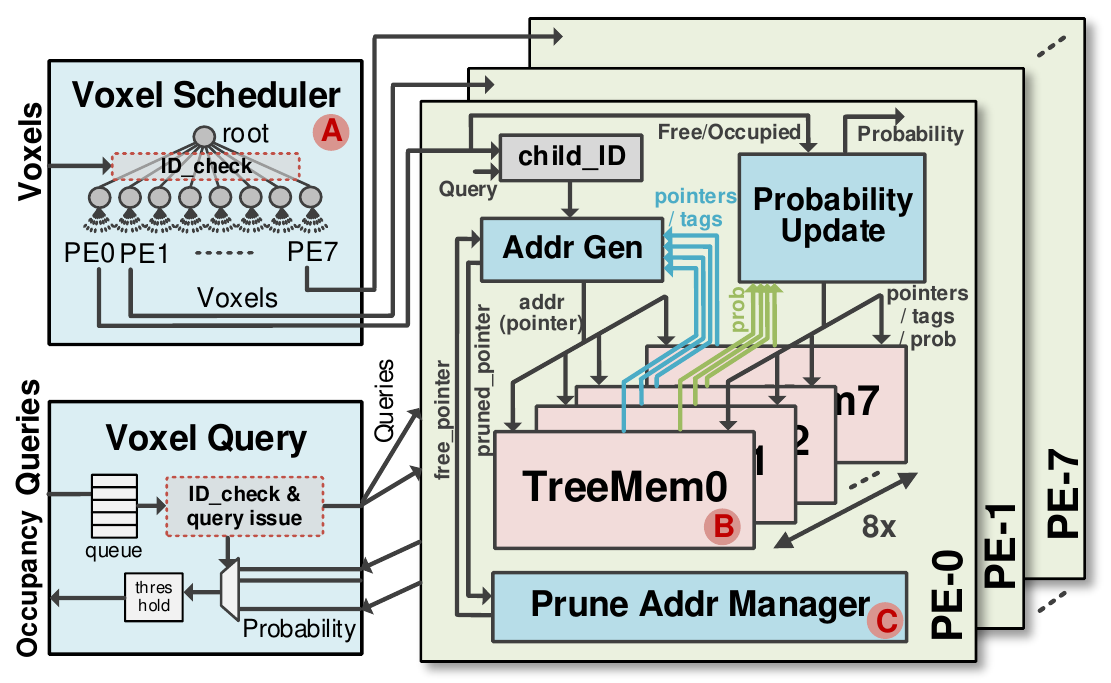}
    \vspace{-20pt}
    \caption{Hardware acceleration techniques for OctoMap.}
    \label{fig:pe_uarch}
    \vspace{-10pt}
\end{figure}

In this work, to parallelize the voxel update while maintaining the update dependency, the entire Octree is partitioned into 8 PE units and store map voxels based on the first-level tree branches. 
As shown in Fig. \ref{fig:pe_uarch}, for each input voxel, it is sent to a voxel scheduler module. Based on the first-level tree branch ID, which is generated by the voxel coordinate, the voxel update request is issued to different PE units. The number of PE is set to 8 (for eight children of each node) to maximum voxel update throughput by up to 8$\times$.

\subsection{Parallel Memory Storage and Efficient Data Structure}

As our workload analysis indicates the voxel pruning is the performance bottleneck, we adopt an 8 parallel bank memory organization in each PE to enables parallel children access, i.e. all eight children can be accessed simultaneously within one clock cycle. This memory structure provides 8$\times$ memory bandwidth improvement and significantly alleviates the node pruning bottleneck, as illustrated in Section \ref{sec:exp}. Together with the PE level data parallelism, the accelerator is able to process 64$\times$ more voxel updates than a single-threaded CPU.

\begin{figure}[!t]
    \centering
    \includegraphics[width=1\columnwidth]{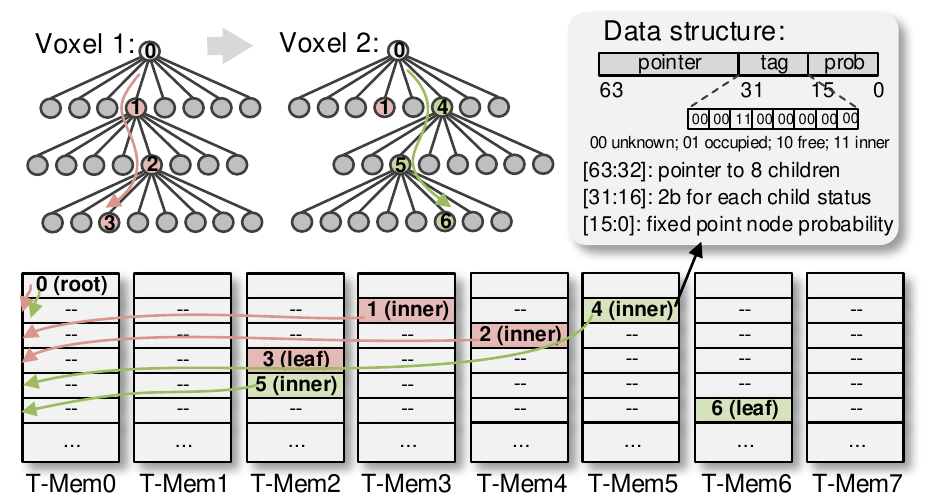}
    \vspace{-20pt}
    \caption{Developed data structure for probabilistic map update.}
    \label{fig:data_structure}
    \vspace{-15pt}
\end{figure}

To efficiently store the voxels in Octree, we introduce a new data storage structure for the probabilistic OctoMap algorithm. This is shown in Fig. \ref{fig:data_structure}. Within each of the 8 memory banks, the data is stored in a 64-bit format, including 32-bits for the memory pointer, 16-bits for the children's status tags, and 16-bits for the fixed-point log-odds voxel probability. 

\textbf{Memory pointer [63:32]:} It is used to point out the memory address of its eight children. The children with the same parent share the same address pointer, while can be distinguished by the memory bank offset (e.g., T-Mem0 stores the children[0]).  

\textbf{Status tag [31:16]:} Each child has a 2 bits status tag. There are 4 possible statuses for each child, i.e., unknown, occupied, free, or inner node (i.e., non-leaf node). 

\textbf{Probability [15:0]:} The occupancy probability of the current node is represented using a fixed-point value. The data type is chosen to have zero loss from the floating-point maps. 

Compared to the previous Octree storage \cite{han2020daduseries}, our work support probabilistic voxel update and multi-voxel access in parallel. Furthermore, the special design data structure also allows easier tree pruning, as it only needs to prune (delete) one memory address pointer for all children, leading to a simple memory address management.

Fig. \ref{fig:data_structure} also provides an illustration example for two voxel updates with a simple tree depth of 3 (the real OctoMap has a tree depth of 16). Based on the input voxel coordinate, we generate the children's address at each tree depth to guide the memory access (i.e., address generation module). For example, the Root node (depth=0) is stored in the first row of the memories, containing the address pointer to its children in depth=1. Hence, Node 1 is updated and stored in the second row of T-Mem 3 because it is the 4th child. Similarly, the leaf Node 3 is stored in T-Mem 2 due to its branch ID. When the tree integrates a new voxel, it will expand the memory storage if the tree branch does not exist before, such as Node 5 and Node 6 in the example. This dynamic memory usage improves overall memory utilization.




\subsection{Dynamic Pruning Address Management}

\begin{figure}[!b]
    \vspace{-20pt}
    \centering
    \includegraphics[width=\columnwidth]{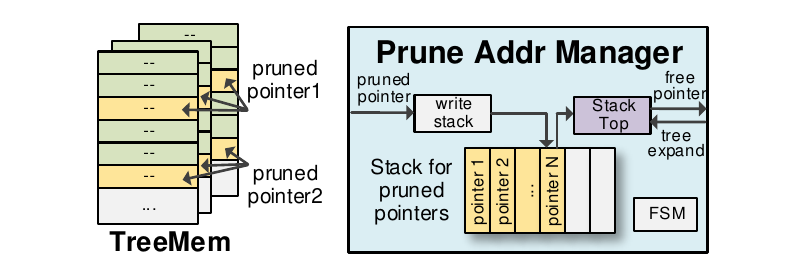}
    \vspace{-25pt}
    \caption{Dynamic prune address manager.}
    \label{fig:prune_addr}
    \vspace{0pt}
\end{figure}

The probabilistic 3D OctoMap has the advantage of dynamic tree pruning to significantly reduce memory storage. 
To efficiently manage the Octree prune and expansion, a prune address manager module is designed in each PE to record the random pruned memory addresses and maximize the memory address reuse. As shown in Fig. \ref{fig:prune_addr}, the prune address manager records the pruned address pointers during tree prune and issues available address pointer during the new tree branch expansion. 
The main logic inside the module is a stack buffer to store all the pruned memory address pointers. 
We use a simple stack buffer instead of a more complex FIFO to manage the dynamic addresses with very small area cost. 

During the tree pruning, the pruned pointer will be written into the stack. At the same time, this prune address manager can provide pruned pointers for new input voxels to reuse the pruned memory space during tree expansion.
Due to this prune address manager, the map memories always maintain a high utilization during the 3D mapping, which also helps to relax the total memory capacity requirement.

%% file: 5_Methology.tex
\section{Overview of \NAME\ Accelerator}  \label{methodology}

\begin{figure}[!t]
    \centering
    \includegraphics[width=0.9\columnwidth]{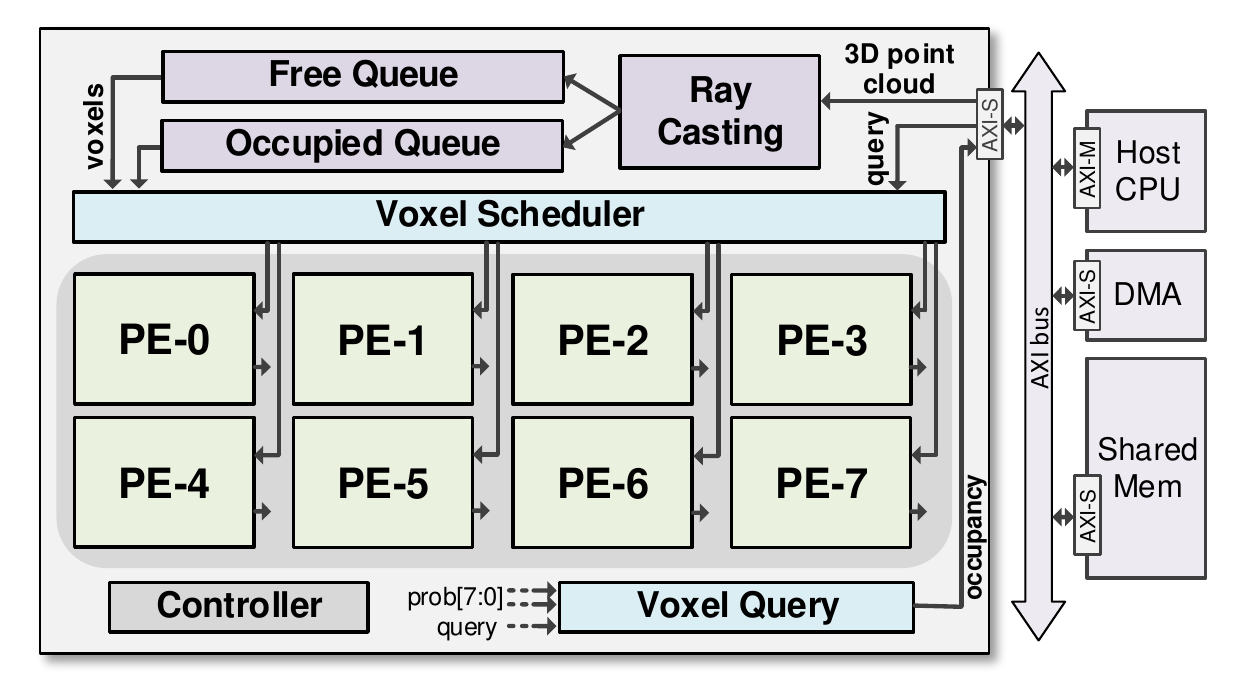}
    \vspace{-5pt}
    \caption{Overview of the proposed \NAME\ accelerator.}
    \label{fig:soc_overview}
    \vspace{-15pt}
\end{figure}

We developed an accelerator solution \NAME to enable real-time 3D OctoMap at the edge. The architecture overview of \NAME\ is shown in Fig. \ref{fig:soc_overview}.
Its key components are the following: 

\textbf{PE Units:} The most essential modules in the accelerator are the PE elements, which store and update the partitioned 3D map. The PE number is set to be 8 to maximize the OctoMap throughput, but it is also scalable depending on the real-time latency and power consumption requirements. In our design, each PE contains 256~kB memory, which consists of 8 32~kB memory banks. The memory size for each PE is chosen based on the consideration of workload sizes and the accelerator area.

\textbf{Voxel Scheduler:} The input voxels for all of the PEs are scheduled using the voxel scheduler. We partition the Octree across the PE units, based on the first-level tree branches. During execution, the voxel scheduler will issue the voxel update task to different PEs based on the voxel coordinates.

\textbf{Ray Casting and Voxel Queues:} The accelerator contains a ray casting module, which performs ray casting operations to identify free voxels between origin and points in the point cloud. The free and occupied voxels identified during the ray casting are stored into queues to be scheduled. The latency of the ray casting has been hidden within the map voxel update.
This module can also be replaced by the more advanced ray casting accelerator \cite{kar2020raycasting} to reduce the voxel update number.

\textbf{Voxel Query:} The accelerator supports a voxel query service, which is a strong requirement for tasks like collision detection in autonomously moving robots. During the voxel query, the probability of the requested voxel is fetched from one PE and sent to the voxel query module. Based on the probability thresholds, the occupancy (i.e., occupied, free, or unknown) of the queried voxel is identified.  

\textbf{Interconnect:} The accelerator is designed with a standard AXI slave interface. A controller module is developed within the accelerator to contain a few sets of configuration registers. To launch the accelerator operation, the user can program the AXI master host CPU to control accelerator configurations. The host CPU also manages to transfer point cloud data from shared memory or DMA DRAM to the accelerator.

%% file: 6_Experiments.tex
\section{Experimental Results}  \label{sec:exp}


\subsection{Methodology} \label{subsec:exp_dse}

We implemented the \NAME\ accelerator using synthesizable SystemC with the aid of hardware components from the open-source MatchLib library \cite{dac2018_hls}. The Verilog RTL is then generated by the Catapult high-level synthesis (HLS) tool. To obtain realistic area and energy results, the RTL of the accelerator is performed using commercial EDA tools based on a commercial 12~nm process. Also, a commercial 12~nm SRAM compiler generates all the SRAM memories. 

\begin{figure}[!t]
    \centering
    \includegraphics[width=0.7\columnwidth]{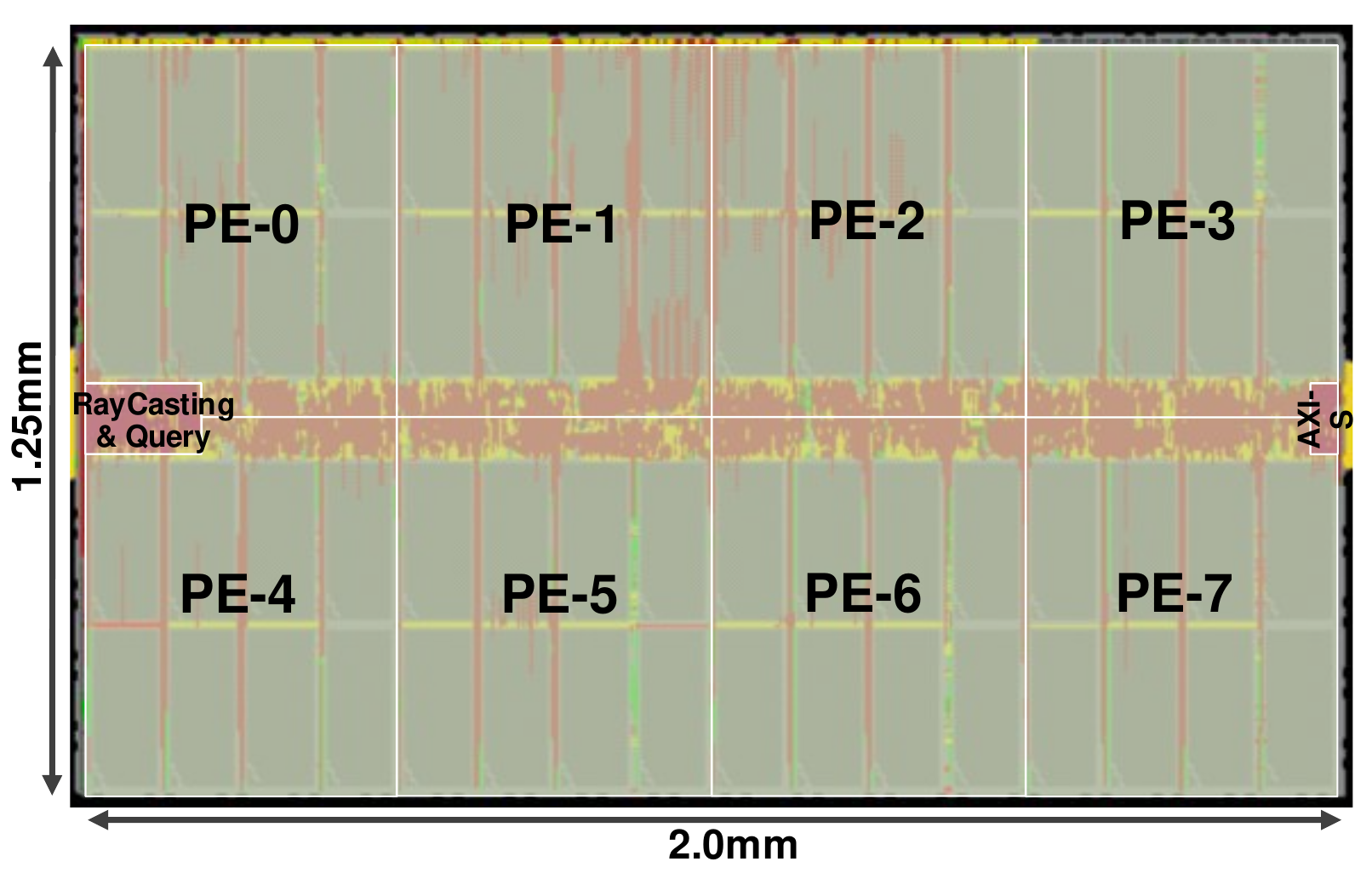}
    \vspace{-5pt}
    \caption{\NAME\ accelerator layout with 8 PEs in 12nm.}
    \label{fig:layout_view}
    \vspace{-10pt}
\end{figure}

The accelerator is signed off at 0.8~V with 1~GHz frequency. We pushed the design to its maximum frequency to obtain better real-time processing performance. The energy, performance, and area results are reported based on the post-P\&R netlist at the typical corner. The layout view of the accelerator with 8-PE (each containing 256kB SRAM to store mappings) is shown in Fig. \ref{fig:layout_view}. The accelerator consumes 2.5 $mm^{2}$ area.

\subsection{Performance Evaluation} \label{subsec:exp_cs}

\begin{figure}[!b]
    \vspace{-20pt}
    \centering
    \includegraphics[width=1\columnwidth]{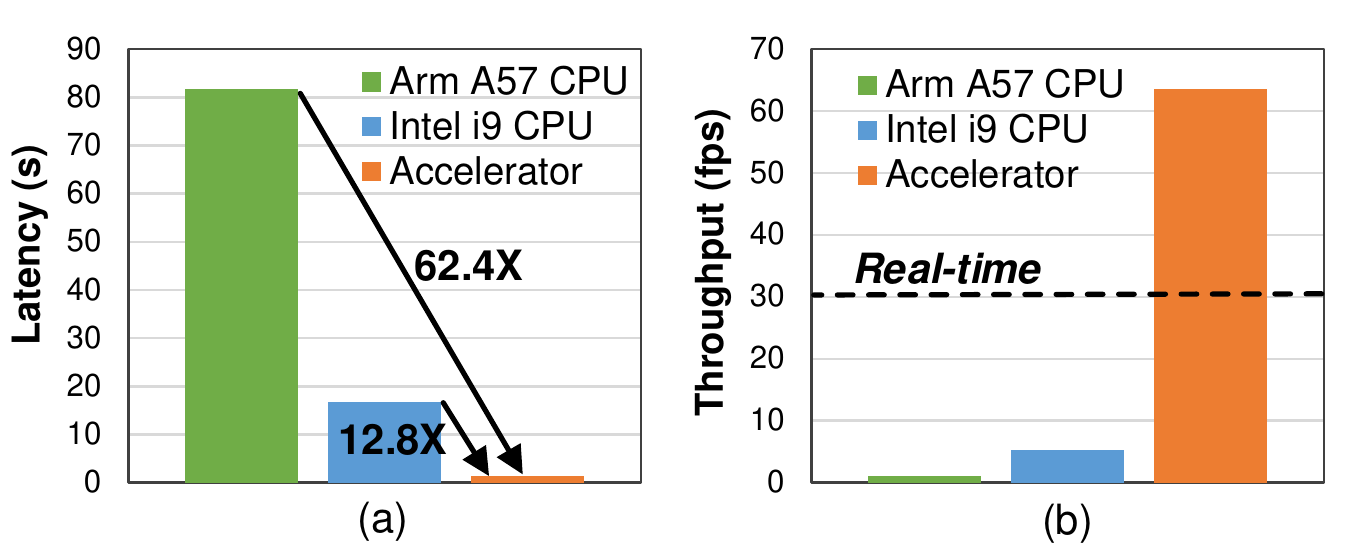}
    \vspace{-15pt}
    \caption{(a) Latency and (b) throughput improvement of the \NAME\ for the 3D map FR-079 corridor.}
    \label{fig:speedup}
    \vspace{0pt}
\end{figure}

We compare the performance of the \NAME\ accelerator with an Intel i9-9940$\times$ desktop CPU and an ARM Cortex-A57 CPU on the Nvidia Jetson TX2 platform. The 3D mappings from the OctoMap 3D scan dataset \cite{octomap_dataset} are used for the evaluations. As shown in Fig. \ref{fig:speedup}, the proposed \NAME\ achieves 12.8$\times$ and 62.4$\times$ latency reduction for the 3D map FR-079 corridor compared to the Intel i9 CPU and the edge Arm A57 CPU. The \NAME accelerator performs 62.4 FPS throughput thanks to the performance improvement, more than 2$\times$ of the real-time application requirement of 30 FPS. 

Table \ref{tab:latency_table} shows the the improvement across different 3D maps. Table \ref{tab:throughput_table} provides the merits of accelerator throughput. Compared to the 5 FPS throughput in Intel i9 CPU, the edge CPU in TX2 can only achieve 1 FPS. In contrast, our accelerator achieves more than 60 FPS for all 3D maps, significantly improving compared to two baseline CPUs. Therefore, with the proposed acceleration techniques, the \NAME\ accelerator can support 3D mapping applications in real-time systems. 

\begin{figure}[!h]
    \vspace{-10pt}
    \centering
    \includegraphics[width=0.8\columnwidth]{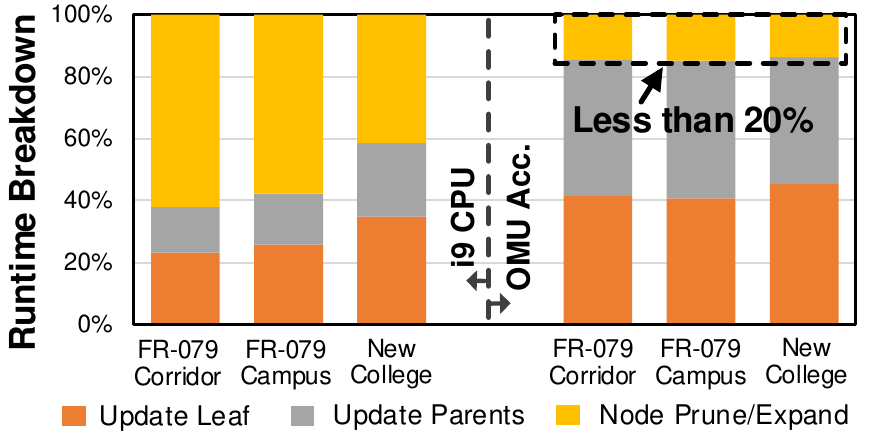}
    \vspace{-10pt}
    \caption{Runtime breakdown in i9 CPU and OMU accelerator.}
    \label{fig:acc_breakdown}
    \vspace{-10pt}
\end{figure}

Fig. \ref{fig:acc_breakdown} shows the breakdown in the accelerator to illustrate the benefits. The Octree node prune and expand only takes less than 20\% runtime in the \NAME\ operations. Compared to the CPU operations, it is evident that the computation bottleneck, i.e., node prune and expand, has been significantly alleviated. This improvement mainly comes from the latency reduction due to the parallel voxel updates and memory storage. Especially, the parallel voxel fetching for all children nodes significantly improves the node update speed and reduces the costly irregular memory accesses in the CPUs.  

\begin{table}[t!]
\centering
\caption{Latency performance (s) comparison table.}
\vspace{-5pt}
\scalebox{0.9}{%
\begin{tabular}{|l|l|l|l|}
\hline
 &   \textbf{FR-079 corridor}  &   \textbf{Freiburg campus}  &   \textbf{New College}   \\ \hline
Intel i9 CPU     &   16.8         & 177.7     &  77.3     \\ \hline
Arm A57 CPU      & 81.7   &897.2    &401.5        \\ \hline
\NAME\ accelerator      & 1.31     &  14.4    &   6.5            \\ \hline
Speedup over i9        & 12.8$\times$     &  12.3$\times$    &   11.9$\times$            \\ \hline
Speedup over A57        & 62.4$\times$     &  62.2$\times$    &   61.7$\times$            \\ \hline
\end{tabular}
}
\label{tab:latency_table}
\vspace{-10pt}
\end{table}%

\begin{table}[t!]
\centering
\caption{Throughput performance (FPS) comparison table.}
\vspace{-5pt}
\scalebox{0.9}{%
\begin{tabular}{|l|l|l|l|}
\hline
 &   \textbf{FR-079 corridor}  &   \textbf{Freiburg campus}  &   \textbf{New College}   \\ \hline
Intel i9 CPU    &   5.23        & 5.03      & 5.04      \\ \hline
Arm A57 CPU      &1.07   &1.0    &0.97        \\ \hline
\NAME\ accelerator       & 63.66     &  62.05    &   60.87            \\ \hline
\end{tabular}
}
\label{tab:throughput_table}
\vspace{-10pt}
\end{table}%

\begin{table}[t!]
\centering
\caption{Energy consumption (J) comparison table}
\vspace{-5pt}
\scalebox{0.9}{%
\begin{tabular}{|l|l|l|l|}
\hline
 &   \textbf{FR-079 corridor}  &   \textbf{Freiburg campus}  &   \textbf{New College}   \\ \hline
Arm A57 CPU      &227.2   &2416.2    &1147.4        \\ \hline
\NAME\ accelerator    & 0.32     &  3.62    &   1.63            \\ \hline
Energy benefit        & 708.8$\times$     &  668.1$\times$    &   703.6$\times$            \\ \hline
\end{tabular}
}
\label{tab:energy_table}
\vspace{-10pt}
\end{table}%

\subsection{Power and Energy Evaluation} \label{subsec:exp_cs}

We analyze the \NAME\ accelerator power after synthesis, place and route using the same 12~nm technology. At 1GHz, the accelerator consumes 250.8~mW power, in which 91\% power is consumed by SRAM access due to the frequency voxel data access and update.
We did not compare the energy consumption of the accelerator with Intel i9 CPU, as the latter is a desktop CPU with a TDP power 165~W. We compare the power and energy with an edge platform (Nvidia Jetson TX2) as it is more representative. During the 3D mapping run on the TX2 platform, the power consumption of Cortex-A57 CPU is recorded, which is between 2.6 and 2.9~Watts. 

Using the average power, we compare the energy consumption between \NAME\ accelerator and the A57 CPU in the Jetson TX2, as shown in Table \ref{tab:energy_table}. We observe that the proposed \NAME\ achieves notably 668$\times$ to 708$\times$ energy efficiency improvement than A57 CPU. Therefore, the proposed \NAME\ provides an efficient and feasible edge computing solution for real-time 3D perception in autonomous machines.




%% file: 7_Conclusion.tex
\section{Conclusion}
3D mapping is an essential process in the perception of autonomous machines. In this paper, we introduce a 3D mapping accelerator to support real-time dense and probabilistic OctoMap efficiently. The proposed \NAME\ hardware accelerator is designed with parallel voxel update and storage, efficient data storage structure, and dynamic prune address management. The accelerator is implemented with a commercial 12~nm technology and evaluated with OctoMap 3D scan dataset. 
Compared to an ARM A57 CPU, the proposed \NAME\ achieves up to 62$\times$ performance improvement and a 63~FPS throughput, enabling 3D mapping in real-time for perception tasks on low-power edge use case deployments.